\documentclass[hidelinks,12pt]{article}

\pdfoutput=1
\usepackage{arxiv}
\usepackage[utf8]{inputenc} 
\usepackage[T1]{fontenc}    
\usepackage{hyperref}       
\usepackage{url}            
\usepackage{booktabs}       
\usepackage{amsfonts}       
\usepackage{nicefrac}       
\usepackage{microtype}      
\usepackage{lipsum}
\usepackage{graphicx}
\usepackage{multirow}
\usepackage{rotating}
\usepackage{subcaption}
\usepackage{color}
\usepackage{pbox}

\title{A Survey of Food Recommenders}

\author{
  Carl Anderson\\
  Weight Watchers International\\
  New York, USA \\
  \texttt{carl.anderson@weightwatchers.com} 
}

\begin{document}
\setcitestyle{square}
\setcitestyle{citesep={,}}
\maketitle

\begin{abstract}
Everyone eats. However, people don’t always know what to eat. They need a little help and inspiration. Consequently, a number of apps, services, and programs have developed recommenders around food. These cover food, meal, recipe, and restaurant recommendations, which are the most common use cases, but also other areas such as substitute ingredients, menus, and diets. The latter is especially important in the area of health and wellness where users have more specific dietary needs and goals. 

In this survey, we review the food recommender literature. We cover the types of systems in terms of their goals and what they are recommending, the datasets and signals that they use to train models, the technical approaches and model types used, as well as some of the system constraints.
\end{abstract}


\section{Introduction}
Food is an important part of everyone’s day and has great social, cultural, and religious significance. However, people can struggle to decide what to eat. First, they may simply be overwhelmed by \textit{choice}~\cite{schwartz2003}. In cities, a hungry couple may be faced with a slew of local restaurants to visit, and many more for online food delivery. A traveler, in town at a conference, wants to explore the best local foods. Where should they go? A cooking enthusiast wants to bake a new cake but which one? There are thousands to choose from! 

Second, people may struggle with \textit{constraints}. A dieter may want help to find tasty and healthy choices. A doctor wants to suggest the best non-repetitive diet to help their patient recover from treatment and its diet-dependent side-effects~\cite{husain2011}. A mother asks, what can I make given that “I have ground beef in the fridge,” or “cherries are currently in season,” or “these lemons are on sale”?   

Given the complexity of the space of ingredients, foods, and restaurants, the multiple use cases, nuanced personal preferences, and the health and religious dietary constraints (Table~\ref{table_summary}), it is not surprising that there many food-related recommenders~\cite{freyne2011,trattner}. In this study, we review the actual and potential use cases (Section~\ref{section_use_cases}), the types of user- and item- data involved (Section~\ref{section_data}), and the technical approaches used or evaluated (Section~\ref{section_technical_approaches}).

\begin{table}[h]
	\caption{High-level summary table that highlights the breadth of food recommender space, covering what is being recommended to whom, how, and why. * represents more speculative examples.}
	\label{table_summary}
	\begin{tabular}{p{45mm}l}
		\hline\noalign{\smallskip}
		Dimension & Examples\\
		\hline\noalign{\smallskip}

		\textbf{Who} are the users? & \textbf{Hungry people}: you might like to order this meal\\
			& \textbf{Cooking enthusiasts}: you might like to make this recipe\\
			& \textbf{Health-conscious}: you'll love this healthy, nutritious lunch\\
			& \textbf{Dieters}: this is a low-calorie but filling and healthy meal\\
			& \textbf{Patients}: doctors suggest that you follow this diet\\		
		\hline\noalign{\smallskip}

		\textbf{What} is being recommended? & \textbf{Ingredient}: you can substitute butter with sour cream for reduced fat and calories\\
			&\textbf{Food}: we think you’ll like these summer rolls\\
			&\textbf{Meal}: we think you’ll like this chicken breast plate with rice and broccoli\\
			&\textbf{Recipe}: try this pecan pie recipe\\
			&\textbf{Recipe collection}: here is a set of salad recipes you’ll love\\
			&\textbf{Restaurant}: you have to try Danny’s Pizza\\
			&\textbf{Cuisine*}: as you like Thai, you might like Indonesian food too\\
			&\textbf{Diet / menu / meal plan}: this is a low-sodium diet that ought to work for you\\
		
		\hline\noalign{\smallskip}
		
		\textbf{When} is it being recommended? & \textbf{Realtime}: where should I eat now; what’s near me?\\
			&\textbf{Batch}: here is your weekly email of recipes, just for you\\
		
		\hline\noalign{\smallskip}

		\multirow{3}{45mm}[\normalbaselineskip]{\textbf{Why} is it being recommended?} & \textbf{Taste}: here is something you might like to eat / make / order\\
			&\textbf{Health \& wellness}: to help people become or remain healthy, to help people lose weight,\\
			&and to help patients recover\\
		
		\hline\noalign{\smallskip}
	   	   
	   \multirow{4}{45mm}[\normalbaselineskip]{\textbf{How} are recommendations generated?} &Collaborative filtering\\
	   &Content-based\\
	   &Hybrid\\
	   &Graphs and clustering\\
	   &Embeddings\\
	   &Self organizing maps\\
	   &Bayesian networks\\
	   &Expert systems, especially case-based reasoning\\
	   
		\hline\noalign{\smallskip}
	   	   
	   \multirow{4}{45mm}[\normalbaselineskip]{\textbf{Where} are recommendations being made?} & Web\\
	   &Mobile, especially apps\\
	   &Text (SMS)\\
	   &Voice*\\
	   
	   \hline
	   
	\end{tabular}
\end{table}

A note on terminology. In this paper, food has a double meaning. It can refer to food in general, as in ``the food industry.'' However, it can also mean a specific food, such as an apple which is a simple food consisting of a single ingredient, or a composite food consisting of multiple ingredients, such as a summer roll. An equivalent term for a food is a dish. The term meal, however, refers to two or more foods, such as a protein and two side dishes, that make up a coherent unit. For instance, on a food delivery site, you may be able to order the ``fried chicken dinner,'' a meal consisting of three foods: fried chicken, cornbread, and macaroni salad.

\section{Use Cases}
\label{section_use_cases}
In this section, we set out some of the actual and potential uses cases that highlight the breadth of what is, or could be, recommended and some of the underlying drivers.

\subsection{Food Delivery}
\label{subsection_food_delivery}
Food delivery is big business—about 1\% of the US food market or \$96B~\cite{hirschberg2016}, consisting of 4.5M U.S. online food orders per day~\cite{chamberlain2015}—and with some apps collating hundreds or thousands of restaurants in a city, the choice can be overwhelming for a user. According the National Restaurant Association, a staggering 46\% of smartphone users in the U.S. use their phones at least once a month to order restaurant takeout or delivery\footnote{\url{https://www.restaurant.org/News-Research/Research/soil}}.

There are at least two major use cases of interest here:

\begin{itemize}

	\item \textbf{Food} or \textbf{meal}: these services recommend specific food or meals to a user, based on their prior ordering history and sense of that user’s taste, to reduce cognitive load. That is, it acts as an information filter, they don’t need to think so hard, and it can help them explore new food (e.g.,~\cite{quiroga2004}). This is the single largest potential use case. While some apps such as Ubereats, Foodler, Halla\footnote{Halla claim that its system "analyzes customer order history...to power restaurant, dish, recipe, and ingredient recommendations, in real-time” \url{https://www.prweb.com/releases/food_tech_innovator_halla_unveils_intelligent_ordering_platform/prweb15658286.htm}} and Caviar learn a user’s taste profile and provide personalized food ordering recommendations, some apps, including some of the major players, simply provide a search experience: what’s popular, near me, and can be delivered quickly. This is likely a wasted opportunity given the user and item data available to those services.

	\item \textbf{Restaurant}: these services recommend restaurants that deliver food. By directing the user to menus that they may wish to explore, this acts as a filter reducing the scope of the meals that the user need consider.

\end{itemize}

One potential use case, but which does not appear to be in the literature, is that of cuisine recommendation\footnote{While various authors predict cuisine from ingredients, e.g.~\cite{jayaraman2017,kumar2016b}, and \url{kaggle.com} have hosted a playground competition to predict cuisine, \url{https://www.kaggle.com/c/whats-cooking}, they are not recommenders.}: given that you like cuisine $X$, you might like similar cuisine $Y$. This could be useful in food delivery apps to broaden the scope of food and restaurants that a user might wish to order from.

\begin{figure}
	\centering
	\fbox{\includegraphics[width=12cm]{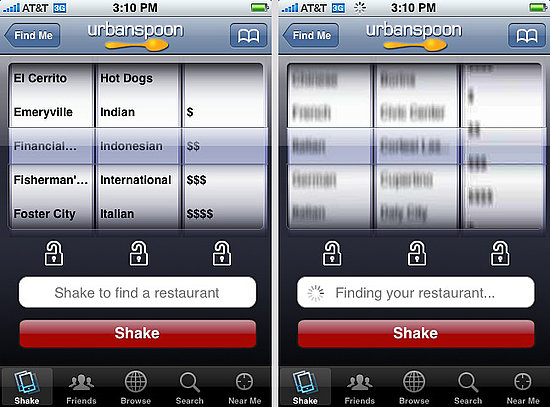}}
	\caption{Urbanspoon’s (now Zomato) provided an innovative interface. The user set a geography, cuisine, and price category then shook the device to generate a random recommendation.} 
	\label{fig_urbanspoon}
\end{figure}

\subsection{Restaurants}
In addition to food delivery from restaurants, recommenders can help users choose which restaurants that they might want to visit in person. These visits might be planned in advance, “where shall we go for dinner next Thursday?” or spontaneous, "I’m ready for dinner now, what are good options near me?”, especially from apps that have access to a user’s location through GPS. These can be valuable for not just locals but for travelers who are new to town. 

Recommendations might be for an individual (“I’m hungry”), or for couples (“What’s a good restaurant for a date?”) or for groups (“Where should we go?”)~\cite{park2008}. 

Urbanspoon developed an app with an early and innovative UI for restaurant recommendations (Figure~\ref{fig_urbanspoon}). Users set a neighborhood, a cuisine and price bracket and then shook the phone to spin the wheels and generate a recommended restaurant.

\subsection{Recipes and Cooking}
Cooking enthusiasts want good recommendations of recipes that they could make. Typically, they’ll want recommendations that take into account their taste but they might also want to constrain the result set by facets such as cuisine, preparation time, number of ingredients, and cooking complexity.

A key constraint-based use case is recommendations based on particular ingredient, such as "what should I cook given that I have chicken thighs in the fridge?” There are several use cases, not just what is in the pantry, but what is available due to seasonality or price.

Cooks might also want recommendations on complementary ingredients, “what goes well with salmon?” or they might want recommendations for health reasons~\cite{oh2010,said}. For instance, a user might want to make a given dish healthier by substituting out an ingredient for one that has a comparable flavor profile but is lower calorie, lower fat etc.~\cite{yan2017}.

Yalvac \textit{et al.}~\cite{yalvac2014} present an interesting but unusual use case: recommending recipes to a group of connected people. For instance, in a co-op or other group-living environment, by understanding what people have bought, and when, they can recommend recipes that can use existing and expiring ingredients among the group thereby reducing the amount of food that needs to be thrown away.

\subsection{Diet and Menus}
Many people have specific dietary restrictions. To some degree, those can be handled through a simple individual filter such as “recommend vegetarian meals.” Other health-based use cases require recommending a set of meals, i.e. a diet (meant in the broadest sense: the kinds of food that a person habitually eats) or weekly meal plan (Reviewed in~\cite{kumar2016a}):

\begin{itemize}
	\item \textbf{Meals}: systems can recommend food based on health requirements, nutrition, and other constraints. What can I eat that is less than 500 calories or less than 12 Weight Watchers SmartPoints\texttrademark? Like food delivery, these systems can help avoid repetition and introduce variety but with a more specific goal in mind such as dietary restrictions, calories, or weight loss, not just taste. A growing interface is voice. Coupled with an appropriate weight loss service, it is not hard to imagine: “Alexa, what do you recommend I have for dinner?”
	
	\item \textbf{Diet and Meal plans}: following a diagnosis, an outcome, or treatment, patients in a healthcare setting might be put on a specific diet. This might set out foods that the patient ought to avoid (e.g. lower salt, fat), ones that they should specifically eat (e.g. high fiber), or overall constraints and targets (total calories per day, weight loss goal). These patients may need help with inspiration or structure to follow a prescribed regimen using some set of allowed ingredients and constraints. Thus, a system that sets out a recommended weekly diet or meal plan may beneficial. This may be true of other groups such as busy parents: “I don’t have time or energy to think and plan each night, recommend a plan for the week.”
\end{itemize}

\begin{figure}[h!]
	\centering
	\fbox{\includegraphics[width=12cm]{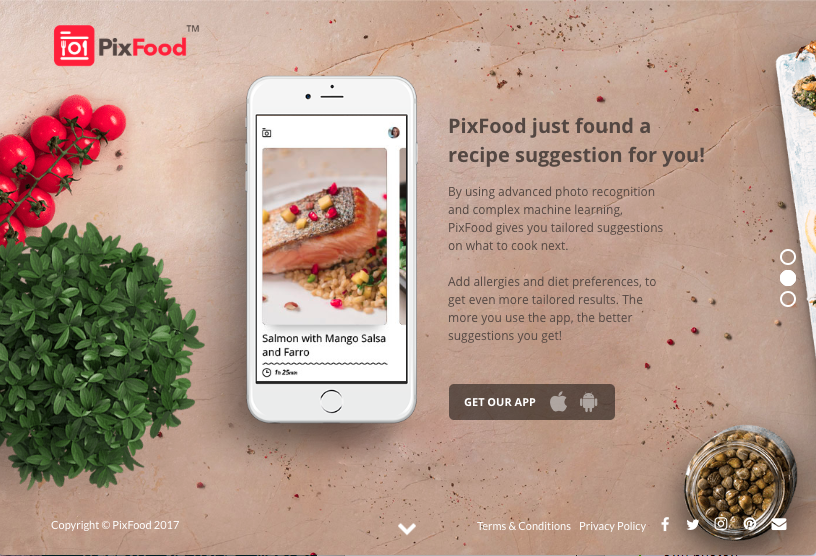}}
	\caption{In pixfood, a user takes a photo of food and receives a recommended recipe.} 
	\label{fig_pixfood}
\end{figure}

\begin{figure}[h]
	\centering
	\fbox{\includegraphics[width=9cm]{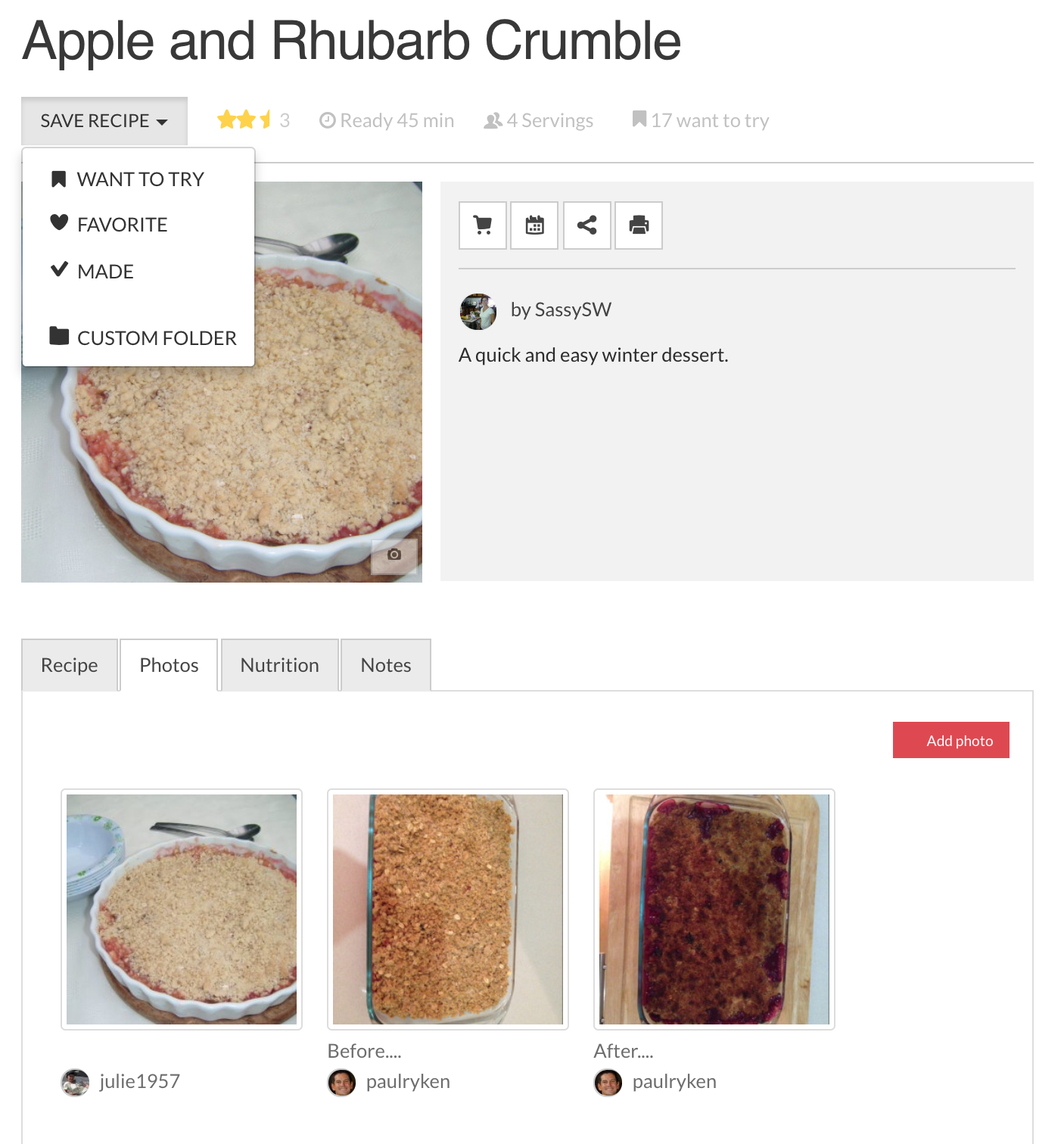}}
	\caption{Example of various signals of interest in a recipe. In this single \url{bigoven.com} recipe, a user can save, rate, favorite, share, indicate that they have made it, that they want to try it, view photos, nutrition, or notes, buy ingredients, add to a meal plan, add to a folder, download, print, and upload their own photo.}
	\label{fig_bigoven}
\end{figure}

\section{Data}
\label{section_data}
In this section, we detail the types of user- and item- data that may be available.

\subsection{User Data}
\begin{figure}[!]
	\centering
	\fbox{\includegraphics[width=6cm]{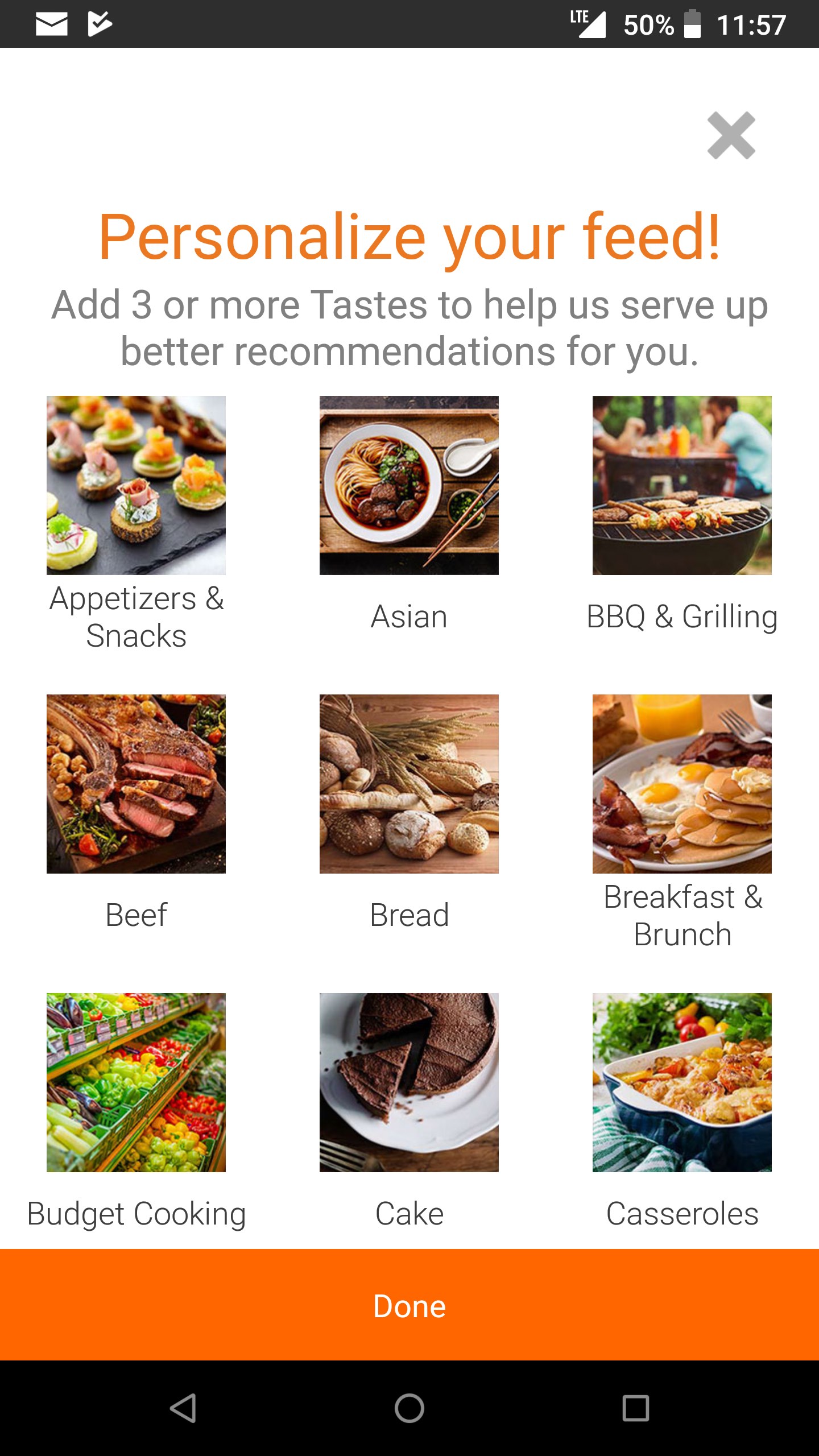}}
	\caption{\url{allrecipes.com}’s dinner spinner app handles cold-start by asking the user to like three or more of 51 different food categories such as cuisine, main ingredient, course, and healthiness.} 
	\label{fig_allrecipes}
\end{figure}

There are many types of potential implicit and explicit user data points that a food recommender can train upon. These vary among systems, but data include:

\begin{itemize}
\item \textbf{Order}: a user orders ingredients or a dish from a food delivery service. Example: Ubereats, Seamless, Caviar, Grubhub, \url{Slice.com}, Amazon Fresh.

\item \textbf{Like}: a user likes (or hearts, gives a thumbs up etc.) an individual meal, recipe, or restaurant. Example: \url{allrecipes.com}, Grubhub.

\item \textbf{Rate}: a user rates an item on a Likert scale. Example: \url{epicurious.com} users rate items on a 4 fork scale.

\item \textbf{Save}: a user saves or bookmarks a recipe, perhaps to their “recipe box.” Example: \url{allrecipes.com}, \url{bigoven.com}.

\item \textbf{Watch}: a user watches a video of a recipe being made. Example: \url{epicurious.com}.

\item \textbf{View}: a user views a set of photos of a dish. Examples: \url{allrecipes.com}, \url{bigoven.com}.

\item \textbf{Post}: for sites with user-generated content, a user posts a recipe, photo of food, a set of nutritional information (\url{loseit.com}, Weight Watchers), or other content for others to discover. In \url{pixfood.com}, the user takes a photo of some food and recipe is recommended (Figure~\ref{fig_pixfood}).

\item \textbf{Share}: for social, community sites, a user can more actively share recipe to followers or members of a group. Example: Figure~\ref{fig_bigoven}.

\item \textbf{Track}: for sites such as Weight Watchers and MyFitnessPal, the act of journaling or tracking food intake might be a key component of the experience.

\item \textbf{Search}: a users searches for food, optionally using a range of filters.

\item \textbf{Comment}: a user comments on a recipe with their feedback or suggested variations and change. Example: \url{epicurious.com}. They may also comment on a social feed, that includes food discussion or images, e.g. Weight Watchers Connect.

\item \textbf{Make}: a user might explicitly signal that they made a recipe with a button such as “I made it!” (e.g. \url{allrecipes.com}; Figure~\ref{fig_bigoven}) or implicitly such as via textual comments (e.g., \url{epicurious.com}).

\item \textbf{Follow}: a user might befriend or follow other users who may share similar tastes. Example: \url{allrecipes.com}.

\item \textbf{Survey}: for onboarding and dealing with cold start, sites might explicitly ask the user to select a set of food-related categories. Example: Figure~\ref{fig_allrecipes}.
\end{itemize}

Figure~\ref{fig_bigoven} shows more than a dozen of these actions and signals in \url{bigoven.com}’s user interface.

\subsection{Item Data}
Typically, a food recommender will require some metadata around the ingredients, foods, and recipes themselves. While one can implement an algorithm such as collaborative filtering without those, such metadata provides the opportunity to use other classes of algorithms, such as content-based recommenders. These metadata are also required for search filters. 

Attributes include:
		
\begin{itemize}
	
	\item \textbf{Allergies}: nut, dairy, wheat etc.

	\item \textbf{Brand}: Stouffers, Del Monte etc. for store-bought branded foods 

	\item \textbf{Cooking time} for recipes.

	\item \textbf{Course}: breakfast, snack, dinner etc.

	\item \textbf{Cuisines}: American, French, Indian etc.

	\item \textbf{Diets}: vegan, vegetarian,  low sugar, low sodium etc.

	\item \textbf{Dish type}: soup, salad etc.

	\item \textbf{Ingredients}: can be used to filter ingredients explicitly in or out for taste, dietary restrictions, or health.

	\item \textbf{Item price, delivery time} and \textbf{price}, and \textbf{open now} for food delivery.
	
	\item \textbf{Number of ingredients} for recipes.

	\item \textbf{Nutrition}: such as calories, grams of protein or fiber, or SmartPoints\textsuperscript{\textregistered}. Some systems simply use “healthy” as an attribute or filter.

	\item \textbf{Preparation time} for recipes.

	\item \textbf{Season}: Thanksgiving, Spring etc.

 	\item \textbf{Tastes}: levels of salt, sweet, bitter, savory, sour, spicy etc.
	
	\item \textbf{Techniques}: grilling, poaching, microwaving etc.
	
\end{itemize}

\subsubsection{Ontologies}
Additional data or metadata that can be useful is a formal food or dish ontology. Ontologies may formalize concepts such as food categories as well as relationships such as dietary restrictions, course types, and preparation methods (e.g.~\cite{bianchini2017,eldosuky2012,kolchin,suksom2010}).

Such ontologies are used to:

\begin{itemize}

	\item \textbf{standardize terms and tags} to provide autocomplete and auto-suggest to a user or to annotate new foods~\cite{bianchini2017}.

	\item \textbf{to query related concepts} such as parent, children, or siblings.

	\item To \textbf{filter} candidate results based on an ontology’s constraints during search.

	\item To aid \textbf{analysis} and aggregation, e.g. to count the number of south-asian dishes in a database where individual dishes might be tagged with lower level concepts such as “Singaporean” and “Thai.”

\end{itemize}

Grubhub developed an ontology with 4000 dishes~\cite{rogers2018}. See also~\cite{hawley2017}. There are a number of open source food-related ontologies, such as BBC Food which describes cooking concepts\footnote{\url{https://www.bbc.co.uk/ontologies/fo}} and FOODON, one version of which contains both food and food production concepts but also comes with a database of 9000 concrete leaf node food products\footnote{\url{http://foodontology.github.io/foodon/}}.

\begin{sidewaystable}

	\caption{Survey of papers in the food recommender spaces. In the users column, "E" means experiment and lists the number of test subjects, while "UC" means a proposed use case. In the purpose column, T = Taste, H = health. \textit{Note: Table continues on next page}.}
	\label{table_bigrefs}
	\renewcommand{\arraystretch}{1.5}
	
	\begin{tabular}{| l | p{20mm} | l | c | p{20mm} | p{20mm}| c | p{25mm} | p{30mm} | }
		
		\hline
		\hline 
		
		
		Reference & \multirow{2}{30mm}{What is being recommended} & Users &  \multirow{2}{25mm}{Purpose\\T: taste; H: health} & \multicolumn{4}{|c|}{Methodology}  & Dataset\\
		\cline{5-8}
		&  &  &  & Collaborative filtering & Content-based & Hybrid & Other & \\		
		\hline
		\hline
		\cite{teng2012} & Ingredients & &  T, H & &  & & Graph & \url{allrecipes.com}\\
		\hline
		\cite{ueda2011} & Ingredients & E: 6 subjects & T & & \centering\textbullet & & & 100 recipes from \url{cookpad.com}\\
		\hline
		\cite{runo2011} & Food & UC: students & T & \centering\textbullet & &&& University canteen menus\\
		\hline
		\cite{suksom2010} & Food & & H &&&& Expert System & ontology only\\
		\hline
		\cite{lo2008} & Food & U: 76 subjects & H & & & & NLP, ontology, latent semantic analysis&\\
		\hline
		\cite{altosaar2017} & Food & & T & & & & Embeddings & \url{allrecipes.com}, \url{epicurious.com}, \url{jamieoliver.com}\\
		\hline
		\cite{zhou} & Food & & T & \centering\textbullet & & & Long short-term memory & Amazon find foods reviews\\
		\hline
		\cite{ravinarayana2016} & Food & & T & & & & Clustering & 197 foods\\
		\hline
		\cite{phanich} & Substitute food & UC: diabetic patients & H &&&& Self-organized map \& clustering & 290 dishes with nutrition from Thai government\\
		\hline
		\cite{tansey2016} & Food, meals & & H & & & & Clustering & \url{loseit.com}\\
		\hline
		\cite{oh2010} & Food, meals & E: 4 subjects & H & & & & Expert system & Book: \textit{Donguibogam}\\
		\hline
		\cite{yan2017} & Food, meals & E: 227 subjects & T & & \centering\textbullet & & VGG CNN & 10k recipes from \url{yummly.com}\\

		\hline 
		\cite{kim2015} & Food, recipes & myfitnesspal users & T &  \centering\textbullet & & & & \url{myfitnesspal.com}\\
		\hline
		\cite{park2008} &  Restaurants & E: 153 subjects &  T & &  & & Bayesian & 20 users + 90 restaurants\\
		\hline
		\cite{sawant2013} & Restaurants & & T &&&&Graph, clustering & \url{yelp.com}\\
		
		\hline\noalign{\smallskip}
		
		\hline\noalign{\smallskip}
		
	\end{tabular}
	\\(continued...)
\end{sidewaystable}

\setcounter{table}{1}
\begin{sidewaystable}
	\caption{(continued). Survey of papers in the food recommender spaces. In the users column, "E" means experiment and lists the number of test subjects, while "UC" means a proposed use case. In the purpose column, T = Taste, H = health.\\\textit{Table continued from previous page.}}
	\label{table_bigrefs_continued}
	\renewcommand{\arraystretch}{1.5}
	
	\begin{tabular}{| l | p{20mm} | l | c | p{20mm} | p{20mm}| c | p{25mm} | p{30mm} | }
		\hline
		\hline 
		Reference & \multirow{2}{30mm}{What is being recommended} & Users &  \multirow{2}{25mm}{Purpose\\T: taste; H: health} & \multicolumn{4}{|c|}{Methodology}  & Dataset\\
		\cline{5-8}
		&  &  &  & Collaborative filtering & Content-based & Hybrid & Other & \\		
		\hline
		\cite{martinez2015} & Restaurants & & T & \centering\textbullet & & & Expert system & Restaurants in Ja\'en province\\
		\hline
		\cite{ramirezgarcia} & Restaurants & E: 50 subjects & T &  \centering\textbullet &  \centering\textbullet & & Fuzzy inference& 60 restaurants in Tijuana\\
		\hline
		\cite{burke2002} & Restaurants & E: 50k user sessions & T &  \centering\textbullet & &  \centering\textbullet & Expert system & EntreeC hosted at \url{kdd.ics.uci.edu}\\
		\hline
		\cite{jagithyala} & Recipes &&T & \centering\textbullet & \centering\textbullet & \centering\textbullet && \url{allrecipes.com}\\
		\hline
		\cite{freyne2010} & Recipes & E: 512 subjects & T & \centering\textbullet & \centering\textbullet & \centering\textbullet & & \url{allrecipes.com}\\
		\hline
		\cite{cohen2014} & Recipes & & T & \centering\textbullet & \centering\textbullet  & \centering\textbullet & & \url{allrecipes.com}\\
		\hline
		\cite{vivke2018} & Recipes & E: 24 subjects & T & \centering\textbullet & \centering\textbullet & & & \url{allrecipes.com}\\
		\hline
		\cite{lin2014} & Recipes & & T & \centering\textbullet & \centering\textbullet & & & \url{food.com}\\
		\hline
		\cite{almedia2015} & Recipes & & T & \centering\textbullet &\centering\textbullet &\centering\textbullet & & \url{food.com}, \url{epicurious.com}\\
		\hline
		\cite{cromwell} & Generated recipes & E: 62 subjects & T, H & &&& Graph & \url{allrecipes.com}\\
		\hline
		\cite{bianchini2017} & Recipes, menus & &  H & & \centering\textbullet & & & \url{bigoven.com}\\
		\hline
		\cite{skeels2018} & Recipe collection & &  T & &  & & Embeddings & Caviar app\\
		\hline
		\cite{husain2011} & Diet & & H & & & & Expert system; Genetic algorithm&\\
		\hline
		\cite{khan2003} & Diet & UC: patients & H & &&& Expert system&CBord hospital system\\
		\hline
		\cite{kovasznai2011} & Diet & UC: patients & H & &&& Expert system&\\ 
		\hline\noalign{\smallskip}
		\hline\noalign{\smallskip}
	\end{tabular}
\end{sidewaystable}

\section{Technical Approaches}
\label{section_technical_approaches}
In this section, we cover the various technical approaches adopted.

\subsection{Collaborative Filtering}
From Table~\ref{table_bigrefs}, collaborative filtering is the most common approach to generating recommendations, often serving as a baseline with which to compare other techniques ~\cite{almedia2015,cohen2014,freyne2010}. This is unsurprising given that it is a standard technique that can work well out of the box without requiring high-quality tags or other metadata for the items. However, its downfall is cold-start for new users and items, e.g.~\cite{jannach}, which some apps counter by surveying the customer for their tastes and interests. For instance, \url{allrecipes.com}’s dinner spinner asks users to rate at least 3 out of 51 categories covering cuisine, main ingredient, course, and healthiness before they can use the app (Figure~\ref{fig_allrecipes}). Similarly, Martinez \textit{et al.}~\cite{martinez2015} require new users to first rate 20 restaurants.

Collaborative filtering is a good approach for sites with a strong community, such as recipe sites in which users follow and rate others’ contributions (e.g., \url{allrecipes.com}, \url{epicurious.com}, and \url{cookpad.com}), and food delivery services where users use ratings to select food and where the sites nudge or incentivize users to provide feedback.

\begin{figure}[h!]
	\begin{subfigure}{.23\textwidth}
		\centering
		\fbox{\includegraphics[width=.95\linewidth]{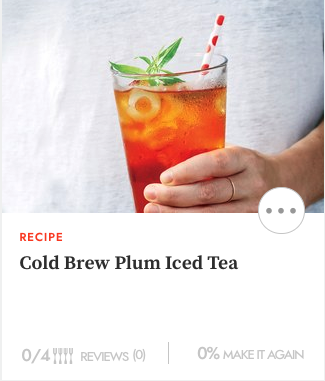}}
		\caption{Liked item}
		\label{fig:sfig1}
	\end{subfigure}%
	\begin{subfigure}{.77\textwidth}
		\centering
		\fbox{\includegraphics[width=.95\linewidth]{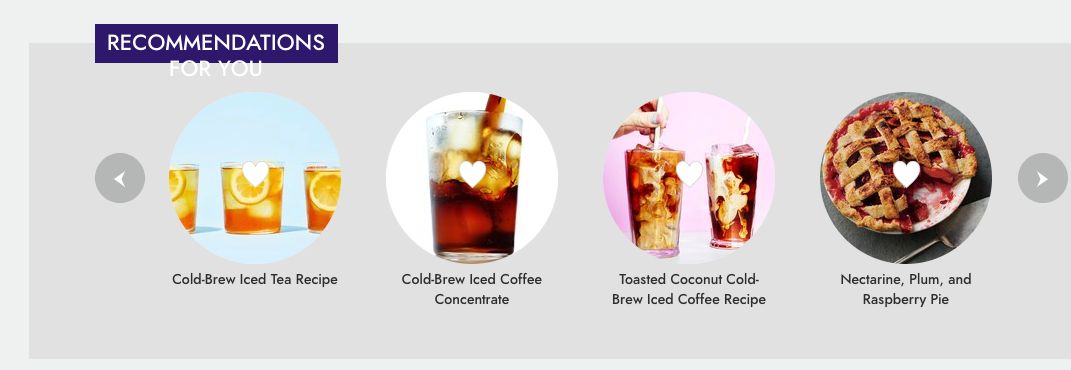}}
		\caption{Four recommendations}
		\label{fig:sfig2}
	\end{subfigure}
	\caption{Example of a (presumably) content-based food recommendation from \url{epicurious.com}. From a clean, cold-start account, liking "Cold brew plum iced tea" (a) leads to the set of four recommended plum and iced-tea based foods (b).}
	\label{fig_epi}
\end{figure}

\subsection{Content Based}
Content-based recommenders are also common (Table~\ref{table_bigrefs}). These offer the potential to pick up on a person’s (literal) taste profile--that they like categories and cuisines such as citrus, Japanese food, medium spice, but not lamb—and serve up similar foods (Figure~\ref{fig_epi}). Such a strategy can potentially learn dietary restrictions from tracked and rated items, provided that there is sufficient item metadata, although that question is perhaps better asked explicitly. You risk offending a user by recommending foods that they cannot eat.

Content-based systems are less likely to serve up false positives, such as foods or categories that you have not shown an interest in, but can lack diversity and serendipity. Such exploration may be especially important for food recommenders in that people don’t like to eat the same things every day, cooks often want to broaden their scope and try new things, and that there is a strong ingredient seasonality at play. Just because you haven’t interacted with strawberry items so far doesn’t show a lack interest, they simply may not have been available at the market until now. (\cite{zhou2010} and~\cite{sheth} provide some approaches to introduce diversity in recommenders, and in~\cite{zhou2010}'s case, without a loss of accuracy.)

In most cases, content attributes apply to the food item as a whole: the cuisine, the nutrition, or the spiciness of the dish. Recipes, however, are an interesting exception. They are a mix of recipe-level attributes as well as ingredient-level attributes. That is, individual ingredients might be associated with a spiciness, nutrition, or might determine the dominant flavor or color. That is, a recipe is a vector of ingredients, each of which might have relevant weights, tags, and attributes. Recipe ratings can also be passed down the ingredients themselves, e.g.~\cite{freyne2010}.

\subsection{Hybrid}
Given the pros and cons of collaborative filtering and content-based recommenders, several studies tried a hybrid approach~\cite{almedia2015,cohen2014,freyne2010} reviewed generally in~\cite{burke2002}. Cohen \textit{et al.}~\cite{cohen2014} assume that users exhibit preference for recipes, but a recipe contains explicit ingredients, each of which has a latent ingredient factor (in 50 dimensions). By factorizing a user-ingredient matrix (rather than user-recipes), they produced a set of meaningful factors that corresponded to groups such as bread, beef, drinks, and barbecue. A user’s profile then becomes a preference across all the ingredients in all the recipes that they have rated, and a candidate recipe is a vector across the ingredient factors, which can be ranked and recommended via cosine similarity.

Freyne and Berkovsky~\cite{freyne2010} built a content-based recommender by passing the ratings of recipes down to the ingredients themselves, weighted equally. However, an ingredient might be part of a set of recipes which received mixed ratings but that ingredient was not the cause of the poor ratings. To counter this, they only considered the positive ratings for ingredients that received mixed ratings. Collaborative filtering is then used to predict a score for unrated ingredients. As such, this hybrid outperformed both their purer collaborative filtering and content-based recommenders.

\subsection{Graphs}
Graphs are another option to capture the relationships among foods which can be used to provide recommendations. Using \url{allrecipes.com} data, Teng \textit{et al.}~\cite{teng2012} constructed two graphs. The first was an ingredient network that captured the co-occurrence of ingredients among recipes. The second leveraged the user-generated suggestions for recipe modifications to create a substitute network. This latter network consisted of communities of functionally equivalent ingredients. As such, they provide a mechanism by which one can suggest healthier substitute ingredients. Moreover, they demonstrated that recipe ratings can be well predicted (accuracy 0.92) by these two networks, with most of the power coming from the ingredient network.

Cromwell \textit{et al.}~\cite{cromwell} extended that work to produce a generative system. By computing network centrality and network community attributes for each ingredient, the system generated 800 salad recipes with 7--9 ingredients. They had a chef make a set of human-generated salads as well as some of these network-generated salads and had human subjects do a blind taste test. While the subjects preferred computer-generated salads less overall, they were rated higher in novelty and the subjects could not always tell which was human- versus computer-generated. 

Ahn \textit{et al.}~\cite{ahn2011} created a flavor network based on the key flavor compounds found in ingredients from gas chromotography. Their analysis allowed them to understand flavor and ingredient combinations and identify authentic ingredient that defined a cuisine (e.g., soy sauce for Asian cuisine).

Together, these graph-based approaches show some early promise in an understanding of flavor profile, ingredient combinations and substitutions, so that a recommender could target a goal, for instance, to make healthier recipes with the same flavor, tailored to a user’s taste profile and history. 

\subsection{Word Embeddings}
Word embeddings extend the vector space model to handle text. While these are not commonly used for food recommenders, there are some interesting examples in the literature. The name of a food, a dish description, or a detailed recipe are all textual documents. Using these inputs, embedding models, such as word2vec~\cite{mikolov2013}, can learn associations among the words and produce not only clusters and associations than can drive recommendations but analogies too. 

\begin{figure}
	\centering
	\includegraphics[width=12cm]{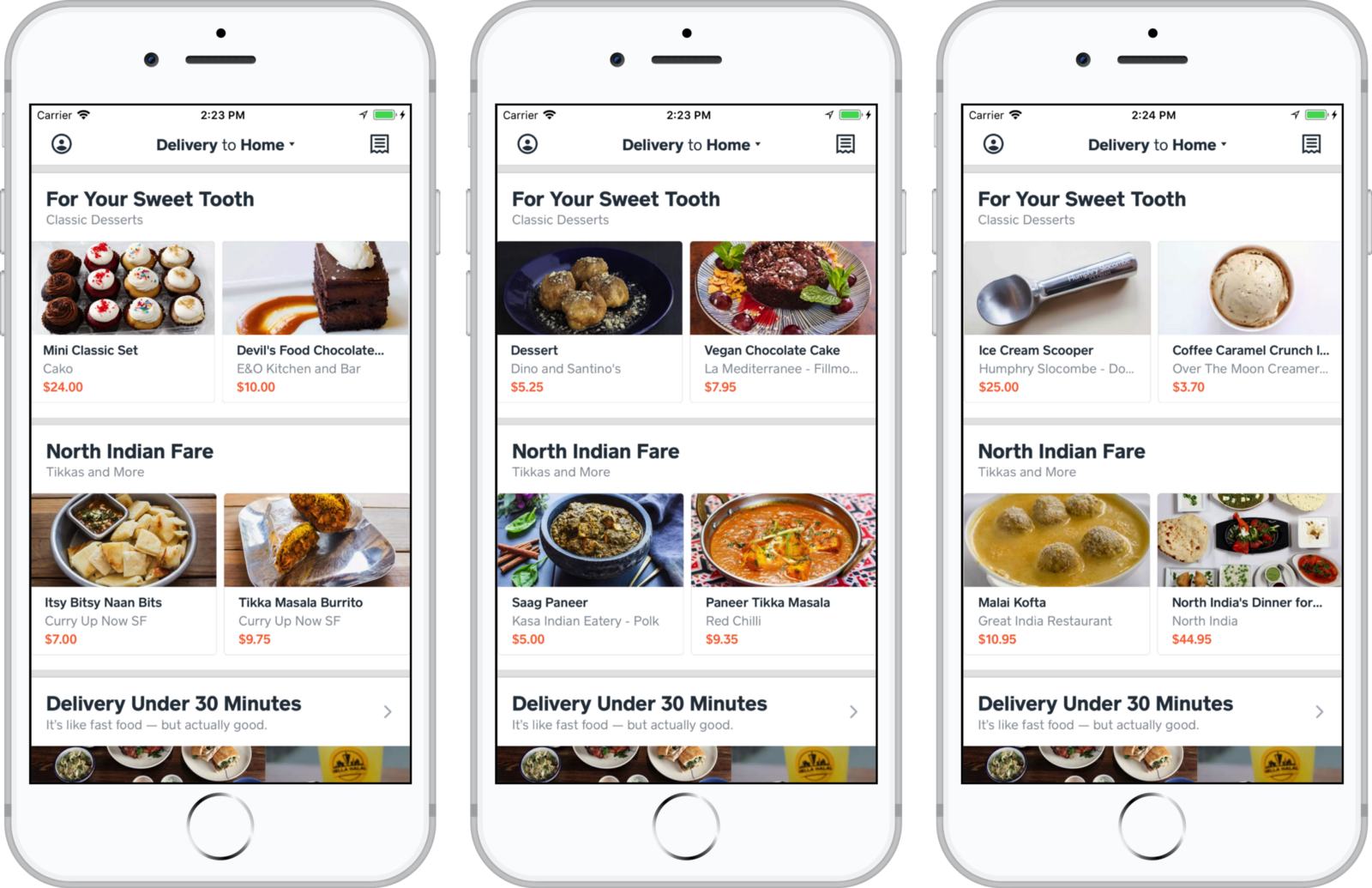}
	\caption{Example of Caviar app’s auto-generated collections. Reproduced with permission from Square (Skeels and Patel~\cite{skeels2018}.} 
	\label{fig_caviar}
\end{figure}

For instance, Skeels and Patel~\cite{skeels2018} used word2vec on restaurant menus, converted menu items into vectors using the same model, averaged the vectors for each word in the menu item phrase, and found the closest category tag out of sets like ``pizza,'' ``dumplings,'' and ``ramen.'' Once the menu items were tagged, they used this to auto-generate collections of items which can then be recommended to users (Figure~\ref{fig_caviar}). Altosaar~\cite{altosaar2017} also used an embedding model but trained on recipe text. After training, the model could complete analogies—food A is to food B as food C is to food D. Given A, B, and C, the system outputs D—with reasonably intuitive responses, such as:

\begin{itemize}
	\item Egg is to bacon as orange juice to \textbf{coffee}.

	\item Bread is to butter as roast beef is to \textbf{sage}.
	
	\item Smoked salmon is to dill as lamb is to \textbf{asparagus}.
\end{itemize}

He suggests that such a system could be used by cooks. For instance, for “lamb, cumin, and tomato,” the system recommended “raisin” as an  ingredient\footnote{See online tool at \url{https://altosaar.github.io/food2vec/\#food-similarity-tool}.}.

Some services, such as \url{loseit.com} and Weight Watchers, allow users to enter and track nutrition. The data consist of a food name and a list of real-valued nutrition values. For example, a 10” burrito tortilla has 5.1g fat, 40g carbohydrates, 2.4g fiber, 2.7g sugars etc. Tansey \textit{et al.}~\cite{tansey2016} use word2vec on the food names to obtain a vector for a food and then concatenated the nutrition data to the vector. They clustered the data to 4000 foods and used another embedding model paragraph2vec~\cite{le2014} to obtain meal vectors. Clustering meals and running paragraph2vec, they obtained diet vectors, which they clustered. Consequently, they were able to produce a multi-level representation of food and nutrition that they argue could be used to recommend meals and food substitutions, to predict user weight loss and churn, and tailoring support messages to user to continue diet tracking.

\subsection{Expert Systems}
Expert systems, or knowledge-based systems, are an older and more traditional model for decision support. In vogue in the 1980s and 1990s, they have the advantage of interpretability and the ability to include explicit knowledge, rules, and overrides from domain experts and experiments, such as clinical trials. Therefore, it makes sense that food recommenders using approaches such as case-based reasoning come from the health domain. A hospital, and their lawyers, are more likely to be comfortable with a system that has explicit human-readable guardrails than say a collaborative filter, which could recommend foods that the patient should not consume. These expert systems can incorporate low-level constraints as well as high-level objectives such as targeted, total nutrition over multiple meals. For instance, Khan and Hoffman~\cite{khan2003} and Kovásznai~\cite{kovasznai2011} both use case-based reasoning with ripple down rules. 

Suksom \textit{et al.}~\cite{suksom2010} uses two ontologies: one is a Taiwanese food ontology consisting of food items in six groups (rice and grains, vegetables, fruits, milk, fats, meat and protein) as well as a personalized food ontology containing personal profile, diet goals, and favorite foods. Given the latter, the recommendation engine produces rules such as

\begin{verbatim}
	if Person.BMI >= 24 
	AND Person.unFavoriteIngredient = Meat 
	THEN Person.RecommendedFood_ingtredientType = (Low_Calorie AND Vegetable). 
\end{verbatim}

These rules can then be used to mine the databases for individual foods that meet those criteria.

\subsection{Other Aspects}
\subsubsection{Clustering}
Like other systems, food recommenders suffer from data sparsity. Not only may the total number of items be large (e.g., \url{allrecipes.com} has about 1 million recipes, discussed later) but items are usually non-standardized. Baby carrots, Julienned carrots, and blanched carrots are all effectively the same item: carrots (e.g.~\cite{lin2014}). Given such variation, the universe of ingredients, food, meals, and restaurants is large, and a random pair of users have rarely experienced or rated the same items. Authors have clustered ingredient~\cite{teng2012},  foods~\cite{phanich,ravinarayana2016,tansey2016}, meals~\cite{tansey2016}, restaurants~\cite{tansey2016}, and users~\cite{tansey2016} using KNN and k-means. This creates larger, denser neighborhoods which improve results. 

Interestingly, Phanich \textit{et al.}~\cite{phanich} used a two-stage process. They constructed and trained a self-organized map (or Kohonen network~\cite{kohonen1982,kohonen2001}) and then clustered those results using k-means. They claim that this provided clusters in which foods within  a cluster had similar amounts of 8 nutrients (protein, fat, vitamin C etc.).

\section{Conclusions}
\label{section_conclusions}
In this review, we covered the recommendation of food, detailing substitute ingredients, food and meals, restaurants, and well as menus and diets (Table~\ref{table_bigrefs} but also see~\cite{bianchini2017,kumar2016a,trattner} for additional papers, especially in the health domain). While we did not find explicit examples of cuisine recommendation, there is such a compelling use case, especially in food delivery and restaurants recommendation, that it likely exists in an app but was overlooked. The closest we found was Skeels and Patel~\cite{skeels2018}. 

These recommenders exist because of the scope and complexity of dishes and meals, the enormous number of possible recipes, and the vagaries of personal preferences~\cite{freyne2010,trattner}. Users need help to navigate this space. There are strong financial incentives to solve these problems and provide a valuable user experience as the markets are large. As mentioned earlier, food delivery is close to a \$100Bn industry in the US with a large number of players. Restaurants are almost \$800Bn in annual sales at 1 million restaurants\footnote{ \url{https://www.restaurant.org/News-Research/Research/Facts-at-a-Glance}.}. Online dieting in the US alone is worth \$1bn\footnote{ \url{https://www.prnewswire.com/news-releases/us-weight-loss-market-worth-66-billion-300573968.html}.}. With a large number of competitors in each of these markets, a good recommender system could sway users to use a particular service.

Researchers use, or have at least tried, numerous approaches to build recommenders~\cite{trattner}. While some are standard out-of-the-box collaborative filtering and content-based algorithms, others have taken some interesting alternative approaches, including latent space embeddings, self-organized maps, and graph-based approaches. (Linear programming, genetic algorithms, and Bayesian methods have also been tried, e.g.~\cite{kovasznai2011,park2008}.) One driver is ingredients in recipes. That is, recipes list individual ingredients and that provides a rich context of ingredient co-occurrence. Also, a recipe details not just ingredients but cooking methods, preparation time and sequencing, cuisine, and other aspects. Thus, there is a rich but complex set of metadata than can be incorporated. 

Deep learning has the potential to be used for recommenders. Yang \textit{et al.} \cite{yan2017} used a convolutional neural net to analyze images of dishes and find visually similar but healthier versions. Serifovic claims to be able to use a convnet to go from an image of food to a recipe (\url{https://github.com/MURGIO/Food-Recipe-CNN}) (See also~\cite{zhou}.) This is likely a growing area to watch for in food recommendation.

One aspect that provides problematic for building food recommenders is the data. While Netflix has less than 10,000 titles (as of 2018; it has actually decreased over time), \url{allrecipes.com} has about 1 million recipes. While this sounds like an advantage (such as mean inter-recipe distance), it means that ratings data per recipe tends to be much more sparse. Lin \textit{et al.}~\cite{lin2014} found an average of 209 ratings per item and 5600 per user on Netflix compared to 38 per item and 4 per user on \url{Food.com}’s 226k recipes. (See Trattner and Elsweiler~\cite{trattner} for discussion of additional challenges for food recommenders.)

This review covered English-language papers only. However, even papers from non-US universities tended to use US-dominated datasets such as \url{allrecipes.com}, Yelp, Yummly, and \url{food.com}. There were some other national, local, and regional datasets (e.g.~\cite{phanich,runo2011,ueda2011}), especially in healthcare but, in general, non-Western datasets were under-represented. This is important because how people perceive, interact, and categorize food varies geographically, nationally, and among religions.

An aspect that was largely missing in the literature was the temporal aspect of eating. People don’t want to eat the same thing every day—although this is more true of dinner than breakfast (Weight Watchers, unpublished data)—but they may have some favorites that they eat frequently. Lin \textit{et al.}~\cite{lin2014} cover some aspects of recency but diet recommenders ought to generate multiple daily menus to counteract patient boredom, but this aspect is largely under-served.

In terms of trends, I suspect that we will increasingly see voice being used an interface, especially in the kitchen. As users get more used to voice interaction, and as home assistants, such as Siri, Alexa, and Google Home, become more common, it will likely become more natural to ask "Alexa, what should I eat for lunch?" This is especially true as kitchen appliances get smarter. Imagine when a fridge knows what food it contains and when it expires. Coupled with a recommender service, it can then suggest what to make. Or, a service can recommend a recipe, consult with the fridge, and order the remaining needed ingredients from a grocery delivery service. Some apps, such as Zopingo already incorporate recommended recipe to grocery delivery. Thus, the notion of a recommended, highly personalized smart meal kit, delivered to your door becomes a real possibility.

\section*{Acknowledgements}
The author would like to thank Nic Chikhani, Mike Skarlinski, Stacie Sherer, and Chris Skeels for valuable suggestions on an earlier draft.

\end{document}